\documentstyle{l-aa}

\newcommand{\mean}[1]{\left\langle #1 \right\rangle}

\input{psfig}

\begin{document}

   \thesaurus{
              04         
              (02.08.1;  
               03.13.4;  
               11.06.1)} 

   \title{An SPH code for galaxy formation problems}

   \subtitle{Presentation of the code}

   \author{John Hultman \and Daniel K{\"a}llander}

   \offprints{J. Hultman}

   \institute{Astronomiska observatoriet i Uppsala,
              Box 515,
              S--751 20 ~Uppsala,
              Sweden\\
              e-mail: pohlman@astro.uu.se}

   \date{Received 22 October 1996 / Accepted 12 February 1997 (A\&A 324, 534)}

   \maketitle
   \begin{abstract}
We present and test a code for two-fluid simulations of galaxy formation,
one of the fluids being collision-less. The hydrodynamical evolution is
solved through the SPH method while gravitational forces are calculated
using a tree method. The code is Lagrangian, and fully adaptive both in
space and time.
A significant fraction gas in simulations of hierarchical galaxy formation
ends up in tight clumps where it is, in terms of computational effort,
very expensive to integrate the SPH equations. Furthermore, this is a
computational waste since these tight gas clumps are typically not
gravitationally resolved. We solve this by merging gas particles in these
regions, thereby limiting the SPH resolution to the gravitational force
resolution, and further speeding up the simulation through the reduction
in the number of particles.
      \keywords{hydrodynamics -- numerical methods -- galaxies: formation}
   \end{abstract}

\section{Introduction}
Recently, there has been a dramatic increase in extra-galactic observational
data through, e.g.,
the Hubble and COBE satellites, and very large ground based telescopes.
These detailed observations have put new demands on theoretical models
of galaxy formation.

Gas dynamics with proper energy dissipation is of fundamental importance
in galaxy formation. Analytic models of
gas dynamics are typically restricted to systems possessing a high degree
of symmetry. Hierarchical galaxy formation, on the other hand, is a highly 
inhomogeneous process. In order to follow the three-dimensional evolution of the
baryonic component in a proto-galaxy, numerical techniques are required.

Smooth particle hydrodynamics (SPH) is a fully Lagrangian numerical method
for gas dynamics. SPH was introduced by Lucy (\cite{Lucy77a}) and Gingold \& Monaghan (\cite{Gingold77a})
to avoid the limitations of grid-based methods. When coupled with a fast
algorithm to calculate gravitational forces, this method has been used
successfully to study hierarchical galaxy formation (Evrard \cite{Evrard88a}, Hernquist \& Katz \cite{Hernquist89a}, Navarro \& White \cite{Navarro93a}).

In this paper we present an SPH implementation, coupled with a gravitational
tree method, that has been specifically developed to study
the formation of galaxies. The code is applied to several test cases
to clarify the advantages and limitations of the implementation.
The code used for the simulations described in this paper is fully Lagrangian,
does not put restrictions on the geometry of the problem, and has a locally
varying resolution both in space and time.

Simulating the formation of galaxies in a hierarchical model is a demanding
task for any numerical method. Small objects form first, and the
characteristic mass of objects then grows rapidly with time, as smaller
objects merge into larger ones. At any given time there is a wide range
of object masses, and the code must be able to handle many different
scales simultaneously. When galaxies merge, part of the gas falls to the
center of the new galaxy. This gas concentration effect is especially
pronounced in simulations
without star formation, where the baryonic mass remains gaseous throughout
the simulation. The gas cores that collect at the center of dark matter
halos are typically to small to be gravitationally resolved, but can still
be the cause of
much of the calculational cost. We avoid this problem by merging
particles in unresolved gas cores, thereby limiting the hydrodynamical
resolution to the gravitational resolution, and furthermore speeding up
the calculations significantly.

\section{The SPH method}

\subsection{SPH principles} \label{sph-basics}

The central concept in SPH is the introduction of the interpolation
procedure.
This is a procedure to define a continuous, 
differentiable function, given values at discrete points in space.
We shall
only outline the procedure, since the method has been presented many times
elsewhere (Hernquist \& Katz \cite{Hernquist89a}, Benz \cite{Benz90a}
and Monaghan \cite{Monaghan92a}).

A common way to smooth a continuous field is to convolve it with a
suitable kernel $W(\vec{x}; h)$,
\begin{equation} \label{kernel-smooth}
  \mean{f(\vec{r})} = \int f(\vec{r}') W(\vec{r} - \vec{r}'; h) d^3r',
\end{equation} 
the integration being over all space. The kernel, or window function,
$W(\vec{x},h)$ goes to zero for large values of $x/h$. the parameter $h$ thus 
specifies the
width of the smoothing window. Furthermore, to preserve the overall
normalization of the field that is smoothed, and to ensure that $\lim_{h\to 0} 
\mean{f(\vec{r})} = f(\vec{r})$, the kernel must satisfy
\begin{equation} \label{kernel-norm}
  \int W(r;h) d^3r = 1.
\end{equation}

In SPH, particles are assigned local values of thermodynamic quantities,
and corresponding continuous fields are defined by convolution of
particle properties with a suitable kernel. The volume integration in
Eq. (\ref{kernel-smooth}) then turns into a sum over all particles
\begin{equation} \label{MC-formula}
  \mean{f(\vec{r})} = \sum_{j=1}^N f(\vec{r}_j) W(|\vec{r} - \vec{r}_j|; h_j)
  \frac{m_j}{\rho_j},
\end{equation}
where N is the total number of particles, $\vec{r}_j$, $m_j$, $h_j$, $\rho_j$
are the position, mass, smoothing length, and density, respectively associated 
with particle $j$. The factor
${m_j / \rho_j}$ is the volume element that SPH associates with particle $j$.

Much of the usefulness of SPH derives from the ease with which gradients
of fields can be calculated. Ignoring surface terms, a direct differentiation
of Eq. (\ref{MC-formula}) yields
\begin{equation} \label{MC-gradient}
  \mean{\nabla f(\vec{r})} = \sum_{j=1}^N f(\vec{r}_j) \nabla W(|\vec{r} - \vec{r}_j|; h_j)
  \frac{m_j}{\rho_j}.
\end{equation}
We see that once the gradient of the chosen kernel has been calculated,
calculations of field gradients proceed similarly to the calculation
of field values. It is especially worth noting, that the sets of particles that contribute to the SPH sums in Eq. (\ref{MC-formula}) and (\ref{MC-gradient})
are identical.

As is common, we use the spline 
kernel, that was first proposed by Monaghan \& Lattanzio (\cite{Monaghan85a}),
\begin{equation} \label{B-spline}
  W(r, h) = \frac{\sigma}{h^3}
  \left\{ \begin{array}{l@{\quad \quad}l}
    \frac{3}{2} (\frac{1}{2} u - 1) u^2 + 1 & 0 \le u < 1 \\
    \frac{1}{4} (2 - u)^3 & 1 \le u < 2 \\
    0 & u \ge 2
  \end{array} \right.
\end{equation}
where $u = r/h$ and $\sigma$ is $1/\pi$ in the three dimensional case.

\subsection{Hydrodynamical equations} \label{hydro-eqns}

The time evolution of the fluid is determined by the usual set of
hydrodynamic equations. Expressed in Lagrangian form,
the SPH expression for the momentum equation has the form
\begin{equation} \label{SPH-pressure}
  \frac{{\rm d} \vec{v}_i}{{\rm d} t} = -\sum_{j=1}^N m_j \left(
  \frac{P_i}{\rho_i^2} + \frac{P_j}{\rho_j^2}  + \Pi_{ij}\right)
  \nabla W_{ij} - \nabla \Phi,
\end{equation}
where $\vec{v}$ is the velocity field and $\Phi$ is the
gravitational potential. $W_{ij}$ is the kernel used for the interaction
between particle $i$ and $j$, and
$\Pi_{ij}$ is an artificial viscosity pressure term, needed to handle shocks
in SPH, described in detail later. For work with galaxy
formation, the gas is usually assumed to be ideal.
The pressure can then be expressed in terms of the density, and the specific
internal energy $u$,
\begin{equation}
  P = (\gamma - 1) \rho u,
\end{equation}
where $\gamma$ is the specific heat ratio, being $5/3$ for a mono-atomic gas,
and the density is obtained through
\begin{equation} \label{SPH-density2}
  \rho_i = \sum_{j=1}^N m_j W(|\vec{r}_i - \vec{r}_j|, h_i).
\end{equation}
The energy equation can be written as

\begin{equation} \label{SPH-energy}
  \frac{{\rm d} u_i}{{\rm d} t} = \sum_{j=1}^N m_j \left(\frac{P_i}{\rho_i^2} +
  \frac{\Pi_{ij}}{2} \right) \vec{v}_{ij} \cdot \nabla W_{ij} +
  \frac{\Gamma - \Lambda}{\rho_i}.
\end{equation}
where $\Gamma$ and $\Lambda$
are non-adiabatic heating and cooling terms, respectively.

In order to get fully symmetric inter-particle forces, 
the kernel term $W_{ij}$, must be symmetric. We have chosen to symmetrize
in the smoothing lengths, so that $W_{ij} = W(r_{ij}, h_{ij})$,
where $r_{ij} = |\vec{r}_i - \vec{r}_j|$ and $h_{ij} = (h_i + h_j)/2$.

In SPH shocks are handled by introducing artificial viscosity to smear
out discontinuities to resolvable scales. We use the artificial viscosity
introduced by Monaghan \& Gingold (\cite{Monaghan83a}) and  Monaghan \& Varnas
(\cite{Monaghan88a}). The viscosity term in Eq. (\ref{SPH-pressure}) is given
by
\begin{equation} \label{vijrij-visc}
  \Pi_{ij} = \frac{1}{\rho_{ij}} \left(-\alpha \mu_{ij} c_{ij} +
  \beta \mu_{ij}^2 \right),
\end{equation}
where $\rho_{ij} = (\rho_i + \rho_j)/2$, and $c_{ij} = (c_i + c_j)/2$ is the
average speed of sound of particles $i$ and $j$. $\mu_{ij}$ is given by
\begin{equation} \label{vijrij-mu}
  \mu_{ij} = 
  \left\{ \begin{array}{l@{\quad \quad}l}
    h_{ij} \frac{ \vec{v}_{ij} \cdot \vec{r}_{ij} }{ r_{ij}^2 + h_{ij}^2 \eta^2 }
   & \vec{v}_{ij} \cdot \vec{r}_{ij}  < 0 \\
    0 & \vec{v}_{ij} \cdot \vec{r}_{ij} \ge 0
  \end{array} \right.,
\end{equation}
where $\vec{v}_{ij} = \vec{v}_i - \vec{v}_j$, $h_{ij} = (h_i + h_j)/2$, and 
$\eta$ is a constant that
prevents singularities for small particle separations. Typically, $\eta = 0.1$,
and $\alpha$ and $\beta$ are set to values close to unity.

There is a disadvantage with this type of artificial viscosity, in that it
introduces shear viscosity into the flow, in addition to the (desired) bulk
viscosity effects. We correct for this as shown in Navarro \& Steinmetz
(\cite{Navarro96a}). All simulations in this paper uses this correction,
except the Collapse test, Sect. \ref{Collapse}, where essentially no shear
flow is present.

The smoothing length for a particle $i$ is set by requiring that the number
of particle neighbors, within a distance of $2h_i$, are roughly equal to a 
constant number, $n_s$.
The smoothing length of a particle is adjusted at each time-step according
to this criterion.
To fulfill this, the smoothing length is updated at the beginning
of each time-step according to
\begin{equation} \label{h-update} 
  h^i = h^{i-1} \frac{1}{2} \left[ 1 +
  \left( \frac{n_s}{n^{i-1}} \right)^{1/3} \right],
\end{equation}
where $h^{i-1}$, $h_i$, and $n^{i-1}$ are the previous and present values of
the smoothing length and the previous number of neighbors respectively.
The number of neighbors is then
calculated using the predicted value of the smoothing length, and if
found to deviate from $n_s$ by more than $n_{tol}$,
the above formula is reiterated. We chose $n_s = 64$ and
$n_{tol} = 10$.

We do not (yet) include the, so called, $\nabla h$ terms, which have been
closely examined by Nelson \& Papaloizou (\cite{Nelson93a}) and Serna et al.
(\cite{Serna95a}), due to the relative complexity of these terms.

\subsection{Gravitational forces and neighbor finding} \label{Gravity}

We have chosen the hierarchical tree approach when calculating gravitational
accelerations. It is a common choice together with SPH,
because, like SPH, it uses no grid and shares a similar kind of flexibility.
Moreover, the gravitational force calculation can then be efficiently
combined with the SPH calculations.

We use the Barnes-Hut algorithm (Barnes \& Hut \cite{Barnes86a}, Hernquist
\cite{Hernquist87a}, and Hernquist \& Katz \cite{Hernquist89a}),
which uses an octal tree to
represent a hierarchical subdivision of space into cubes.
We have introduced a small modification in the tree
construction, in that cells are recursively subdivided until the sub-cells
contain eight or less particles. (The standard Barnes-Hut algorithm
continues subdividing space until a cube contains one or zero particles.)
This modification significantly reduces the number of cells in the tree,
and has the additional advantage that particles can be placed on top of each
other. The latter is sometimes useful when setting up initial conditions
for simulations that involve both a gas and a dark matter component.

When computing the acceleration or potential of a particle,
we use the standard Barnes-Hut criteria. The tree is traversed, root cell first,
level by level, and a cell is accepted as a force term if $s/d < \theta$,
where $s$ is the size of the cell and $d$ is the distance to the center of
mass of the cell. $\theta$ is usually in the range $0.5 - 1.2$.
Rejected cells are sub-divided further. We avoid self-forces by rejecting
cells that the particle itself resides in. (Salmon \& Warren \cite{Salmon93a}).
Quadrupole moments are optionally included.

The tree construction, as well as the traversal is fully vectorized.
The ``breath-first'' approach was preferred over the ``depth-first''
method, because it leads to better vectorization with individual
particle time-steps (Makino \cite{Makino90a}).

The gravitational softening is implemented with kernel softening
(Gingold \& Monaghan \cite{Gingold77a} and Hernquist \& Katz 
\cite{Hernquist89a}), where the force between two  particles
is calculated by considering one particle to have a radial density
distribution given by the same spline kernel as for
the SPH interactions, Eq. (\ref{B-spline}).

Gravitational softening parameters $\epsilon_i$ can be specified for
each particle. Typically,
the softening will be different for the dark matter component and the gas
component, and sometime regions with different resolution are used. The
softening is chosen so as to keep two\/-body relaxation at reasonable
levels. Typically, we calculate the two-body relaxation time-scale
(Binney \& Tremaine \cite{Binney87a}) for gravitationally bound objects as
\begin{equation} \label{relax-time}
  t_{relax} = \frac{N R}{8 \mean{v} \log \Lambda},
\end{equation}
where $N$, $R$ and $\mean{v}$ are, respectively, the total number of particles,
radius and mean velocity, for the object under consideration.
$\log \Lambda = \log (b_{max}/b_{min})$, is known as the {\em Coulomb
logarithm\/}. We use impact parameters, $b_{max} = R N^{-1/3}$, according to
Smith (\cite{Smith92a}). $b_{min} = \epsilon_{min}$, the minimum softening
parameter in the object.

The need to symmetrize pairwise forces occurs also here, and for
particle-particle interactions we symmetrize the
softening lengths, $\epsilon_{ij} = (\epsilon_i + \epsilon_j)/2$,
in the same way as the SPH smoothing lengths. The particle-cell
interactions are not straight-forward to symmetrize in a hierarchical tree
method, and we do not do this.

Instead of starting from scratch, we begin the neighbor search
where the tree traversal for the gravitational force stops, making the
calculations more efficient.
The gravity-neighbor calculations typically
constitutes over 60\% of the computing time. The gravity-neighbor package is
calculating for one particle at a time, and can therefore be executed in
parallel. It was straight forward to implement this on shared memory (SMP)
architectures, like the CRAY Y-MP. Generally this reduces calculation times by
a factor of two.

\subsection{Time integration} \label{time_integr}

We use the second order Runge-Kutta (RK) integrator, with individual
time-steps, described by Navarro and White (Navarro \& White \cite{Navarro93a}).
The main reason for this choice is that it allows the time integration errors
to be estimated, and thereby gives good control over particle time-steps.
There is no need to directly incorporate any stability criteria, e.g.,
the Courant condition.

The system is always advanced with the smallest time-step, and
a subroutine keeps track of which time-levels of particles are
at either the beginning of a time-step, or at the middle of a time-step, and
need to have derivatives of dynamical variables calculated. A linked list
structure is used for the bookkeeping of the particles different time levels.
The overhead caused by this book-keeping constitutes less than 1\% of the
computing time for a typical simulation.

We use an implicit scheme for the integration of the energy equation,
similar to Hernquist \& Katz (\cite{Hernquist89a}) and Anzer et al.
(\cite{Anzer87a}). A trapezoidal formula
\begin{equation}
  \label{trapez} u_i^{n+1} = u_i^n
    + (\dot{u}_i^{n+1} + \dot{u}_i^n) \Delta t_i /2,
\end{equation}
is solved iteratively, together
with  Eq. (\ref{SPH-energy}), at each individual time-step $\Delta t_i$.
A combined Newton-Raphson and bisection
scheme, as described in Press et al. (\cite{Press86a}), is used to find
the root of the energy equation (\ref{trapez}). In absence of cooling and
heating terms, Eq. (\ref{SPH-energy}) is linear in $u_i$, and Eq.
(\ref{trapez}) is solved in one iteration cycle.

When strong radiative cooling is present, the integration tolerance
criterion on the internal energy equation would become very prohibitive,
regardless of what type of criterion is used, (Courant e.t.c.).
Physically, this occurs when the cooling
time-scale becomes much shorter than the local dynamical time-scale
\begin{equation}
  t_{dyn} = \sqrt{\frac{3 \pi}{4 G \rho_i}}.
\end{equation}
It would be extremely expensive in terms of computing time to integrate the system at the time-step imposed by the cooling time-scale. Moreover, for the
purpose of tracking the dynamical evolution of the system, it is not
necessary to let the time-step become much shorter than the local dynamical
time-scale. Therefore, the accuracy constraint on the energy equation is not
allowed to limit the time-step below $10^{-2} \cdot t_{dyn}$.

Although the implicit scheme is unconditionally stable, there will still
be errors in the time integration. We have found that a damping of the cooling terms, similar to Katz \& Gunn (\cite{Katz91c}), improves the accuracy of the
time integration, without  affecting the dynamical evolution.
A particle is not allowed to lose more than half its internal energy during
one time-step, by cooling. We use a sharp cutoff on the cooling part of the
derivative. This has the effect of broadening the fast time variation into a
resolved exponential decay over a few time-steps. The reason why this
works is that it occurs only in regions where the cooling time-scale is
much shorter than the dynamical time-scale. The
energy that is not radiated away in one time-step, will be radiated away in the
following few, while the other variables, like density and velocity,
can be considered to be roughly constant.

Typical radiative cooling functions often have multiple peaks. We have noted
this may cause multiple roots in the solution of the energy equation. Of course,
only one root is correct, i.e., the one closest to the root at the previous
time-step. We solve this problem by finding all roots and accepting
only the closest one. This may seem both computationally costly (and
mathematically impossible, but the cooling functions used, are tabulated at a
finite number of points. Typically, 128-256, and therefore these calculations
take insignificant computational time.

\section{Merging}

\subsection{Merging of particles} \label{Merging}

Our code was written with simulations of hierarchical galaxy formation
in mind.
A typical feature of such simulations is extreme clumping of matter.
The gas collapses into very dense gravitationally bound
objects, where cooling time scales are much less than the local dynamical 
time-scale.
Some of these objects contain very little angular momentum, and the gas
therefore continues to contract until the contraction is artificially halted by
the finite numerical force resolution. This clumping leads to two problems.

First, the suppression of gas contraction on unresolved scales is completely
artificial, and also puts a limit on the gravitational binding energy.
This makes it easier for gas to be ejected from bound objects during violent
mergers, and could lead to an underestimate of the fraction of matter that
has collapsed to galactic densities at a given time.

Second, this is a challenge for SPH because accurate time integration
of the gas in extreme high density regions requires large amounts of
computational work. We have chosen to set the SPH smoothing lengths by
requiring a roughly constant number of interacting neighbors. This leads
to very small smoothing lengths in regions with extremely high density.
To maintain stability, it is therefore
necessary to use very small time-steps in these regions. We have found
that the computational cost for evolving gas in gravitationally unresolved
cores of compact objects can dominate the total run-time. This is a very
undesirable situation.

The problems with small time-steps in unresolved regions can be somewhat 
alleviated, by requiring that the SPH smoothing always be larger than
the gravitational smoothing. This limits the maximum gas resolution to
roughly the gravitational force resolution. However, the problem then occurs
in the SPH force evaluations instead. Setting a lower limit on the SPH
smoothing in high density regions leads to large numbers of interacting
neighbors. This makes the force evaluations very expensive, due both to
the large amounts of terms in the SPH loops and the cost of finding the
SPH neighbors. The code deteriorates towards an inefficient implementation
of direct summation. Limiting the hydrodynamical resolution in this way
actually slows the code down.

The reason that the run-time can increase when the resolution is decreased,
by setting a lower limit on the SPH smoothing length, is that the accuracy
of the SPH force calculations increases. 
A lower limit on the SPH smoothing length leads to large numbers of 
hydrodynamically interacting neighbors in unresolved regions, and therefore
reduced sampling errors in the SPH force calculations.
An alternative would be to only use a random subset of the interacting 
neighbors in the SPH calculations. This is
consistent with the SPH formalism, but it introduces an extra source
of noise in forces.
It would also be ineffective to first find all neighbors of a particle, and then
discard most of them only to use a small subset in the actual force
evaluation.
A tree search can be implemented to find a pseudo-random subset of a particles
neighbors, one by one, by using a depth-first search. This would make it
unnecessary to find large numbers of neighbors. It
looks non-trivial to vectorize such a scheme,
and we have found that large subsets have to
be used for good accuracy. 

We avoid all these problems by not imposing any lower limit on $h$, thus ensuring
efficient force evaluations, and by instead merging particles in high density 
regions,
thus avoiding small time-steps and further speeding up the simulation by
reducing the number of particles. This has previously been tried with
success by Monaghan \& Varnas (\cite{Monaghan88a}) for simulations of cloud collapse, and by Steinmetz (\cite{Steinmetz96a}) for galaxy formation problems.

Merging of particles in high density regions can
keep the hydrodynamical resolution from exceeding the gravitational resolution,
without slowing down the force evaluations, thus greatly speeding up simulations
without sacrificing accuracy. 

The first step is to identify high density regions that are severely affected
by gravitational smoothing.
In these regions it is pointless to calculate substructure on scales smaller
than the gravitational smoothing length.
We therefore merge particles in these regions into more massive, and fewer,
particles.
The dynamics on these small, unresolved, scales cannot be followed in
our code, with or without merging of particles.
The dynamical effects of particle merging should be small outside the regions
where merging occurs.
To satisfy this requirement all global dynamical quantities must be left
unchanged by the merging process.
Thus, when merging two particles into one, global quantities like mass, kinetic 
energy, internal energy, potential energy and angular momentum must be preserved.

To make sure that global dynamical quantities are accurately conserved,
the following five conditions have to be met before two particles are allowed
to merge.
($A_1$, $A_2$, etc. are tolerance parameters.
The quantities $m_i$, $\rho_i$, $\vec{v}_i$, $\epsilon_i$ and
$\phi_i$ are the mass,
local gas density, velocity, gravitational softening length and
gravitational potential of particle $i$.
A ``cm'' subscript indicates a quantity that is evaluated in the center of mass
of the two merging particles. $\vec{r}_{ij} = \vec{r}_i - \vec{r}_j$ is
the relative position of the two merging particles.)

I) The two particles are in a region where gravitational forces are severely
affected by limited force resolution.
\begin{equation} \label{merge-crit-1a}
  \min{(\rho_i, \rho_j)} > A_2 \max{(\rho_{{\epsilon}_i}, \rho_{{\epsilon}_j})}
\end{equation}
where
\begin{equation} \label{merge-crit-1b}
  \rho_{{\epsilon}_i} = \frac{3 n_s m_i}{32 \pi {\epsilon}_i^3},
\end{equation}
which is the density that would be calculated for particle $i$ if
all particles were evenly distributed within radius $2 \epsilon_i$, and they all
had mass $m_i$. In regions where the density is much higher than $\rho_{\epsilon}$,
the mean inter-particle distance will be much less than $\epsilon$, and
gravitational forces will be affected by the limited resolution.
Using $n_s m_i$, instead of summing up the mass of all neighbors,
makes the density criterion stricter for more massive particles, in regions
with particles of varying masses. This evens out spatial particle mass
fluctuations, by letting less massive particles merge before more massive ones.

II) The two particles are SPH neighbors and within one smoothing
length of each other.

The particles should be close together to minimize the artificial displacement
of gas properties that occurs when two particles are merged.

III) The resultant particle mass of the merger is below the preset limit
\begin{equation} \label{merge-crit-3}
  m_i < A_1 m_{initial},
\end{equation}
where $m_{initial}$ is the gas particle mass before any merging has taken
place.
This is a limit on the amount of merging that can occur in a simulation,
and can be useful as a guarantee that the gas resolution does not fall below
a preset limit.

IV) The angular momentum lost is small.
\begin{equation} \label{merge-crit-4}
  \left|\vec{r}_{ij} \times \left(\vec{p}_{i_{cm}} - 
                                    \vec{p}_{j_{cm}}\right)\right|
  < A_3 L_{scale}
\end{equation}
where $L_{scale}$ is a relevant angular momentum scale for the problem at
hand, which in all cases discussed here is equal to the total angular momentum
of the system.

V) The kinetic energy lost is small compared to the gravitational potential
energy,
\begin{equation} \label{merge-crit-5}
   \frac{m_i m_j}{m_i + m_j} (\vec{v}_{i} - \vec{v}_{j})^2 < 
   A_4 (m_i + m_j)\phi_i.
\end{equation}

VI) The particles are synchronized in time and have just completed a full
time-step. This is necessary to avoid breaking the time integration scheme.

These criteria are ordered so that the least computationally expensive ones
are checked first. The computational cost for the whole merging procedure
is then negligible.
Whenever criteria I-VI are met the two particles merge, and form a new
particle.
The new particle is given properties to conserve mass, momentum and thermal
energy. Furthermore, we let the gravitational smoothing be proportional to 
$(m_i + m_j)^{\frac{1}{3}}$. Changing gravitational smoothing changes the
potential. If conservation of binding energy is desired, it is safer to leave
the gravitational smoothing unchanged. The reason is that the internal energy
have to be changed and this may cause unphysical decrease in entropy.

When two particles merge, the resolution in that region changes.
This is accompanied by a small change in the SPH smoothing
length in the region.
To avoid large discontinuous changes in SPH quantities, which could lead to
spurious effects in the time integration scheme and could set up artificial
shock waves, a given mass element is not allowed to merge more than once per
local dynamical time.
This ensures that particle merging is a smooth process where SPH quantities
have time to adjust to changes in the resolution. We have found this criterion
to be crucial to accurately reproduce the results of simulations without
particle merging.

Throughout this paper the values of the tolerance parameters are
$A_1 = 10, A_2 = A_3 = A_4 = 10^{-4}$. With these values we have always
found that
energy and angular momentum are conserved to well within one per cent.
It should be noted, this choice of parameters allow for rather aggressive
merging, sometimes making more than half the gas particles to merge, but all
merging occurs in unresolved cores, and it does work well in the tests
presented here.

\section{Tests}

\subsection{Test of the gravitational interaction; King models.} \label{BH-test}

For objects like galaxies, which have a tremendous number of constituent bodies,
the dynamics may be described by the Boltzmann equation.
A dark matter component can usually be considered collision-less in
contrast to the collisional gas component, for which the Boltzmann equation
can be well approximated with the hydrodynamical equations. In an N-body method
discreetness effects are unavoidable, due to the finite number of particles used.
A practical way to test an N-body code, is to run it on a steady
state system. Conservation of energy, and linear and angular momentum, can
be examined. Also important is that the effects of two-body relaxation
may be investigated.
For a steady state system the potential is time-independent, and
thus the energies of individual particles should be conserved. 
Moreover, if the system
is spherically symmetric, individual particle angular momenta are
conserved. Due to the motions of the particles in an N-body model the
potential will never be strictly time-independent, the forces not strictly
central, and there will always be some
relaxation. The particles energies and angular momenta will diffuse, and this
effect can be measured.

The most practical static models to choose for testing purposes
are King models (King \cite{King66a}), since they are of finite
extent in phase space. Following Binney \& Tremaine (\cite{Binney87a}), King models have a phase space density

\begin{equation} \label{fk-distr}
  f_K({\cal E}) =
  \left\{ \begin{array}{l@{\quad \quad}l}
	\rho_1 (2 \pi \sigma^2)^{-\frac{3}{2}} (e^{{\cal E} / \sigma^2} - 1)
	& {\cal E} > 0; \\ 0 & {\cal E} \leq  0,
  \end{array} \right.
\end{equation}
where the relative energy, $\cal E$, and the relative potential,
$\Psi$, are defined as

\begin{equation}
  {\cal E} \equiv -E + \Phi_0 = \Psi - \frac{1}{2} v^2 \quad\mbox{and}\quad
  \Psi \equiv - \Phi + \Phi_0 .
\end{equation}

The parameter $\sigma$ is a measure of the one dimensional velocity dispersion,
and $\rho_1$ is a mass-normalization constant.
Integrating over velocity space we get the density as a function of $\Psi$,

\begin{equation}
  \rho_K(\Psi) = \rho_1 \left[ e^{\Psi / \sigma^2}
  erf \left( \frac{\sqrt\Psi}{\sigma} \right) - 
  \sqrt{ \frac{4 \Psi}{\pi \sigma^2}} \left( 1 +
  \frac{2 \Psi}{3 \sigma^2} \right) \right]
\end{equation}
where $erf(x)$ is the error function. $\Psi$ will satisfy the Poisson equation,

\begin{equation}
  \frac{{\rm d} }{{\rm d} r} \left( r^2 \frac{{\rm d}  \Psi}{{\rm d} r} \right) = -4 \pi G r^2 \rho_K(\Psi),
\end{equation}
which is an ordinary differential equation for $\Psi(r)$ and can be integrated
numerically once we have suitable boundary conditions. For positive $\cal E$,
which is our region of interest, we must have positive $\Psi$, thus the value
of $\Psi(0)$ is one condition. We note that the enclosed mass inside radius
$r$ can be given in terms of $\Psi$ as

\begin{equation}
  M(r) = - \frac{r^2}{G} \frac{{\rm d}  \Psi}{{\rm d} r}.
\end{equation}

Restricting ourselves to a finite central density, we see that $(d \Psi / dr)
 = 0$ at $r = 0$ is a natural condition. Since $\Psi(0)$ is positive and
$(d \Psi / dr)$ is negative for $r > 0$, $\Psi(r)$ must decrease and become zero
for some radius $r_t$ where also the density will vanish. This radius is known
as the {\em tidal radius}.

King models are commonly parameterized in terms of $\Psi(0) / \sigma^2$, because
the dispersion parameter $\sigma$ just specifies the velocity scale. The mass
scale is still free, but can be specified with a change of scale, i.e., by
adjusting the parameter $\rho_1$.
The radial Poisson equation is solved numerically using a Runge Kutta
method. The enclosed mass $M(r)$, is calculated in order to distribute
the particles.
Particles are initially placed on a grid in a spherical region and then the grid
is stretched into the desired density profile.

In general, it can be hazardous to use grid-setups for very cool collision-less 
systems, but this will not be the case here since the particles are
started with substantial velocities that are randomly distributed.
The advantage with the grid placement, over a random placement,
is that random fluctuations are minimized, and it is
easier to compare the realization with the analytic potential. The system
will also be closer to equilibrium.

\begin{figure}
  \psfig{file=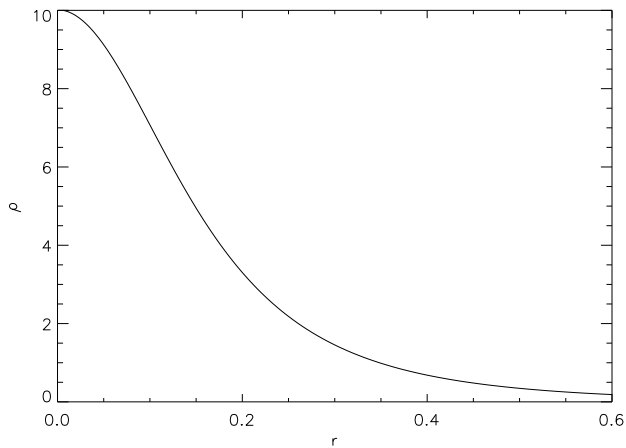,width=8.8cm}
  \caption{\label{kingdens_fig} Density profile of a King model with central
				potential $\Psi(0) = 5 \sigma^2$. Units are
				given by $G = M = R = 1$.}
\end{figure}

Particles are given random velocity directions, isotropically distributed.
Random velocity magnitudes are then assigned to the particles, with a
distribution given by Eq. (\ref{fk-distr}).

The velocity probability density is given by

\begin{equation}
  f_V(v) = f_K(v|r) \equiv \frac{f_K(v,r)}{f_R(r)}
\end{equation}
where $f_K(v|r)$ is the conditional probability density and

\begin{equation}
  f_R(r) \equiv \int f_K(r,v) d^3 v = \rho_K(r).
\end{equation}
The distribution function is then
\begin{equation}
  F_V(v) = 4 \pi \int\limits_0^v f_K(v'|r) v'^2 dv',
\end{equation}
and can may integrated in closed form.

In order to be able to compare results with the tests of Hernquist \& Barnes
(\cite{Hernquist90a}) and Huang et al. (\cite{Huang93a}) we set
up a King model with central potential $\Psi(0) = 5 \sigma^2$. Units were
chosen such that the gravitational constant $G = 1$, the total mass $M = 1$ and
dispersion parameter
$\sigma = 0.762$. This gives a total three-dimensional velocity dispersion of
unity, and thus a total energy of $-1/2$. The tidal radius is $2.18$.

\begin{figure}
  \psfig{file=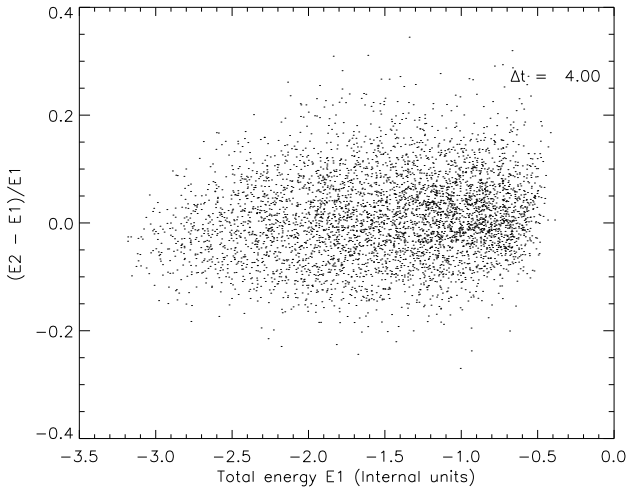,width=8.8cm}
  \caption{\label{Ediff0008} The relative change in individual particle
				energies, as a function of the energies,
				Measured at times $t = 4$ and $t = 8$.
				Internal units are given by $G = M = R = 1$.}
\end{figure}

Four different runs were made with $N = 4626$ or $N = 15238$. A softening
parameter $\epsilon = 0.025$ was used in all runs, and they were continued
to time $t = 8$, which corresponds to around 20 half-mass dynamical times.
The Barnes-Hut parameter $\theta$ was 1.0 in all runs, except run 3, where
$\theta = 0.1$ was used. All runs used individual time-steps, except run 4, where
a constant time step $\Delta t = 1/48$, was used.
In order to test the relaxation effects we examined the relative changes in
particle energies

\begin{equation}
  x_i \equiv (E_i(t_1) - E_i(t_2))/E_i(t_2),
\end{equation}
between times $t_1$ and $t_2$, where $E_i = v_i^2/2 + \Phi_i$
In Fig. \ref{Ediff0008}, the results for run 1 is shown. In Table
\ref{bh_tab}, the standard deviation $S$
and mean value $\bar{x}$ are presented for the runs. Absolute deviations were
also examined,
but they had the same behavior as the standard deviations.
The changes in energies are
expected to be a random walk diffusion process because of small random
accelerations, caused by the random noise in the particle forces. The
diffusion is expected to behave such that

\begin{equation} \label{diffusion}
  S^2 \approx D_{S} \Delta t,
\end{equation}
where $D_{S}$ is an empirical diffusion constant. $D_{S}$ was
calculated with a linear least squares fit of $\log S$ as function of
$\log \Delta t$. The diffusion rates are slightly larger, but comparable to
those of Hernquist \& Barnes (\cite{Hernquist90a}).

\begin{table*}
  \caption{ \label{bh_tab}
		Data for the various King-model runs. Relative changes in
		particle energies, between times $t = 4$ and $t = 8$.
		Conserved quantities.}
  \begin{flushleft}
    \begin{tabular}{llllllllllll}
      \hline\noalign{\smallskip}
      Run & $N$ & $\theta$ & $\Delta t$ & $S$ & $\bar{x}$ &
 	$D_{S}$ & $\Delta E/E$ & $\Delta L/L$ & $\Delta R_{CM}/r_t$ &
	$\Delta V_{CM}/V$ & CPU \\
      \noalign{\smallskip}\hline
      \noalign{\smallskip}
      1 & 4626 & 1.0 & 0.0039* & 0.078 & $-3.3 \times 10^{-3}$ &
	0.035 & $5.4 \times 10^{-3}$ & $4.8 \times 10^{-4}$ & 0.011 & 0.036 &
	1.94 \\
      2 & 15238 & 1.0 & 0.0063* & 0.041 & $-7.8 \times 10^{-4}$ &
	0.020 & $5.5 \times 10^{-3}$ & $1.4 \times 10^{-4}$ &
	$8.1 \times 10^{-3}$ & 0.026 & 9.39 \\
      3 & 4626 & 0.1 & 0.0039* & 0.080 & $-5.4 \times 10^{-3}$ &
	0.036 & $5.6 \times 10^{-3}$ & 0.016 & $2.3 \times 10^{-5}$ &
	$7.3 \times 10^{-5}$ & 32.1 \\
      4 & 4626 & 1.0 & 0.020 & 0.076 & -0.011 &
	0.036 & 0.021 & $7.2 \times 10^{-4}$ & $5.8 \times 10^{-3}$ &
	0.019 & 0.83 \\
      \noalign{\smallskip}\hline
      \noalign{\smallskip}
    \end{tabular}
  \end{flushleft}
  * shortest individual time step. Scales $L$ and $V$ are of order unity. CPU-	times are in hours on a HP 735 workstation.
  \smallskip
\end{table*}

\begin{figure}
  \psfig{file=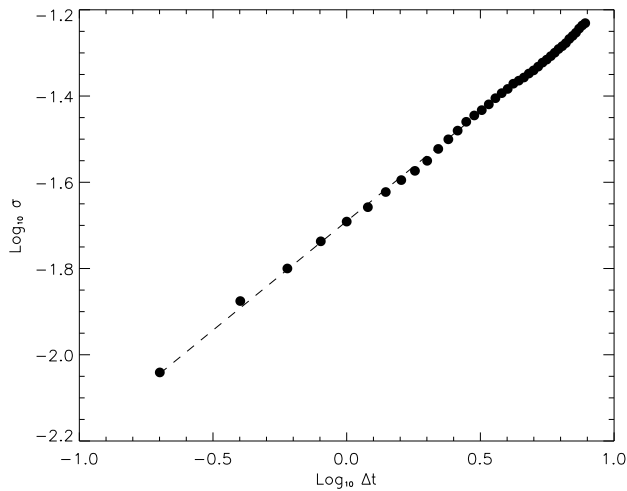,width=8.8cm}
  \caption{\label{Edev0039} The standard deviation of relative particle 
                                energies, $S$, as a
			 	function of the time, for run 2.
				The slope of the dashed line is 0.506.}
\end{figure}

Conservation of energy, linear and angular momentum, and
center of mass position was also examined. The maximum deviations during each
run, are listed in Table \ref{bh_tab}. The time evolution of $S$ is
plotted in Fig. \ref{Edev0039}. The straight line implies a fairly good fit of
Eq. (\ref{diffusion}).
In agreement with what was found by Hernquist \& Barnes (\cite{Hernquist90a}), 
the only parameter having a significant
effect on the diffusion rate was the number of particles used.

Judging from the CPU time used, run 4 seems to be over two times as fast as run
1, but run 4 uses a constant time step, five times larger than the smallest time
step of run 1. Run 1 conserves energy roughly a factor of four better than
run 4.
Three time levels are occupied for run 1, and the benefit of individual
time-steps is roughly a factor of two.

\subsection{The collapse test} \label{Collapse}

One of the most common tests for astrophysical SPH codes is the collapse
test of Evrard (\cite{Evrard88a}). It has
also been presented by Hernquist \& Katz (\cite{Hernquist89a}),
Steinmetz \& M{\"u}ller (\cite{Steinmetz93a}), Nelson \& Papaloizou 
(\cite{Nelson94a}),
and Serna et al. (\cite{Serna95a}). One dimensional, (spherically
symmetric), finite difference solutions have been
calculated by Thomas (\cite{Thomas87a}) and Steinmetz \& M{\"u}ller (\cite{Steinmetz93a}). We will henceforth refer
to this test as the ``collapse test''. The reason why this test is popular is
that it is simple to set up, and that it tests an ``SPH + gravity'' code on
the aspects of adiabatic flow, shocks and gravitational collapse.
The initial setup for this problem is a
gas sphere, of mass $M$ and radius $R$, with density profile

\begin{equation} \label{1/r-dens}
  \rho(r) = \frac{M}{2 \pi R^2} \frac{1}{r}.
\end{equation}

The gas is initially isothermal with an internal energy $u = 0.05 GM/R$, and the
velocity is zero everywhere. Other test parameters are the specific heat ratio
$\gamma = 5/3$, the artificial viscosity parameters $\alpha = 1$, $\beta = 2$
and $\eta = 0.1$, the Barnes-Hut tolerance for the gravitational interaction
$\theta = 0.8$ and a softening parameter $\epsilon = 0.1$. The number
of particles used was $N = 3828$, and the number of neighbors in the SPH
summations was $n_s = 64$.

\begin{figure}
  \psfig{file=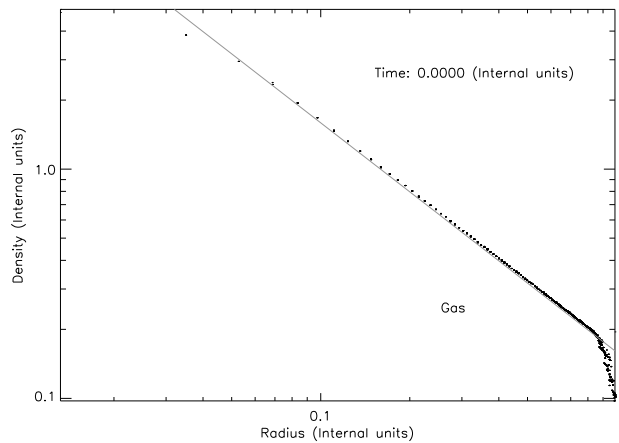,width=8.8cm}
  \caption{\label{cllps_dens_fig} The initial $1/r$-density profile for the
	collapse test. The dotted line is a plot of the desired density,
	Eq. (\ref{1/r-dens}). The SPH density of the particles follows
	the desired density, except at the very center $r < 0.04$, and
	at the outer edge $r > 0.9$, where boundary effects are visible.
	Internal units are given by $G = M = R = 1$.}
\end{figure}

We use the same kind of stretched grid set up, as in Sect. \ref{BH-test}. It
gives a more relaxed initial configuration than the
corresponding random distribution. For this particular distribution
the errors in the SPH-density are very small, as can be seen in Fig.
\ref{cllps_dens_fig}. 

\begin{figure*}
  \psfig{file=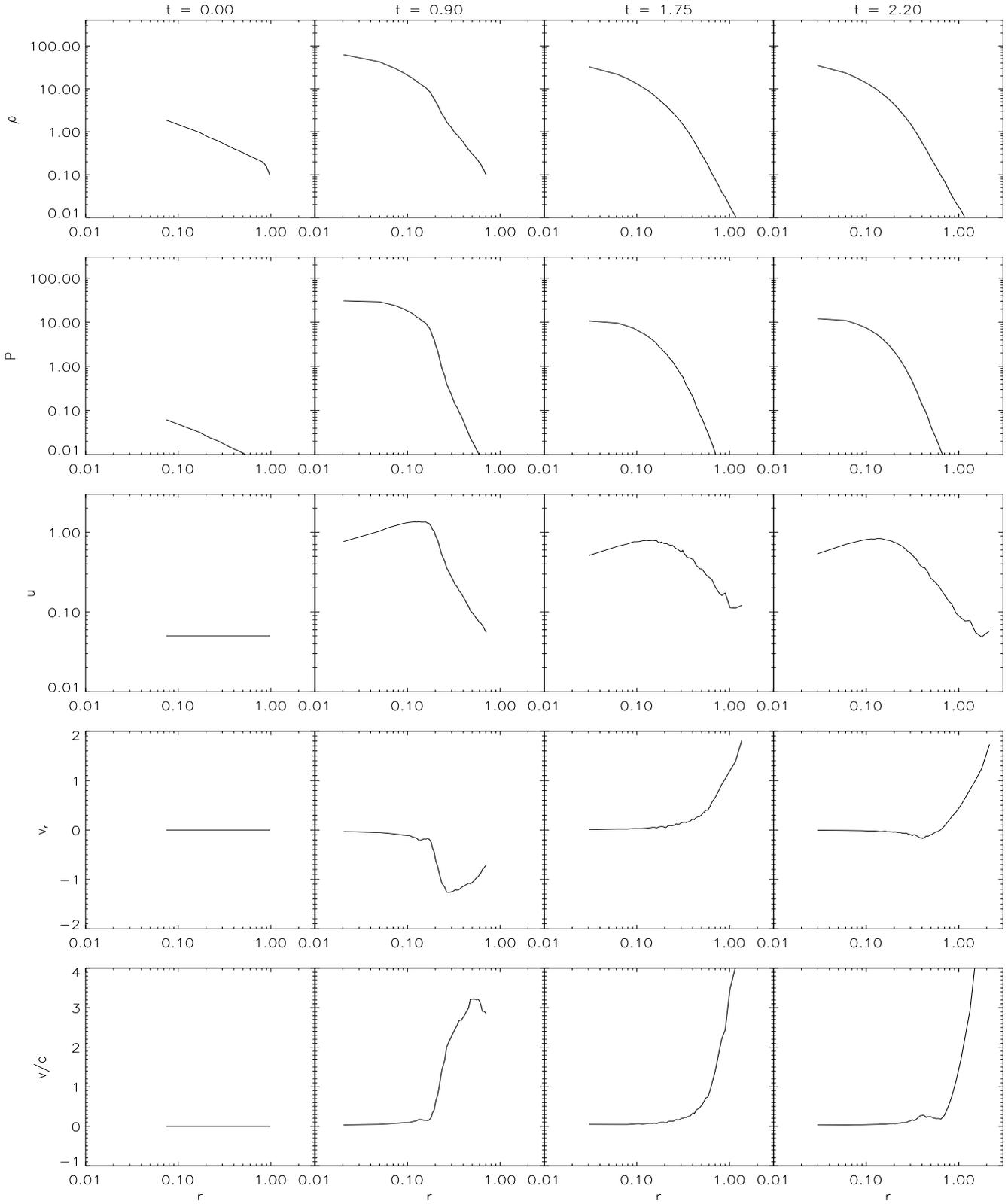,width=18cm,height=22cm}
  \caption{\label{cllps_fig} Density, pressure, internal energy, velocity and
		Mach number, for some different times, in the ``collapse test''.
		Units are given by $G = M = R = 1$}
\end{figure*}

A good way to check for SPH sampling errors, is to perform an SPH evaluation
of a constant unity field, putting $f(\vec{r}_j) \equiv 1$ in Eq.
(\ref{MC-formula}).
It is important that no edges etc., where the sampling necessarily is bad due to the neglect of surface terms are within the investigated volume.
Correspondingly, the gradient of the unit field may be examined.
The statistics
of the SPH representations of a unit field and its gradient, for the initial
configuration in Fig. \ref{cllps_dens_fig}, are presented in Table \ref{one_tab}. Although the result
for a unit field shows very little spread for this particular configuration, the
result for the corresponding gradient field shows considerably more spread.
However, these fluctuations are much smaller than those corresponding to
a random placement setup, as shown in Table \ref{one_tabR}. To avoid boundary
effects, when calculating the numbers in Table \ref{cllps_dens_fig} and
Table \ref{one_tabR}, only a selection of 1800 particles within radius 0.7
were used. 

\begin{table}
  \caption{ \label{one_tab}
		SPH sampling errors for a $1/r$-density profile,
		using a stretched 
  grid setup.}
  \begin{flushleft}
    \begin{tabular}{lllll}
      \hline\noalign{\smallskip}
      quantity & mean & max & min & $\sigma$ \\
      \noalign{\smallskip}\hline
      \noalign{\smallskip}
      unit field  & 1.000 & 1.002 & 0.9989 & $4.495 \times 10^{-4}$ \\
      $ | \nabla({\rm unit field}) | $ & 0.07042 & 0.1953 & 0.01107 & 0.03604 \\
      \noalign{\smallskip}\hline
      \noalign{\smallskip}
    \end{tabular}
  \end{flushleft}
\end{table}


\begin{table}
  \caption{ \label{one_tabR}
		SPH sampling errors for a $1/r$-density profile,
		using random 
  placement.}
  \begin{flushleft}
    \begin{tabular}{lllll}
      \hline\noalign{\smallskip}
      quantity & mean & max & min & $\sigma$ \\
      \noalign{\smallskip}\hline
      \noalign{\smallskip}
      unit field  & 1.008 & 1.352 & 0.6744 & 0.1024 \\
      $ | \nabla({\rm unit field}) | $ & 2.898 & 9.989 & 0.2110 & 1.3079 \\
      \noalign{\smallskip}\hline
      \noalign{\smallskip}
    \end{tabular}
  \end{flushleft}
\end{table}


Units in the collapse test are chosen so that $G = M = R = 1$, and the
usual plots of density,
pressure, internal energy, velocity and Mach number are shown in Fig.
\ref{cllps_fig}. The results show good agreement with what have been 
reported by the authors mentioned above.

\begin{table}
  \caption{ \label{coll_tab}
		Conserved quantities for the collapse test.}
  \begin{flushleft}
    \begin{tabular}{lllll}
      \hline\noalign{\smallskip}
      $N$ & $\Delta E/E$ & $\Delta L/L$ & $\Delta R_{CM}/R$ & CPU \\
      \noalign{\smallskip}\hline
      \noalign{\smallskip}
      3828 & $9.7 \times 10^{-3}$ & $1.0 \times 10^{-4}$ &
	$3.8 \times 10^{-4}$ & 5.25 \\
      \noalign{\smallskip}\hline
      \noalign{\smallskip}
    \end{tabular}
  \end{flushleft}
  Scale $L$ are of order unity. CPU-time are in hours on an HP 735 workstation.
  \smallskip
\end{table}


\begin{figure}
  \psfig{file=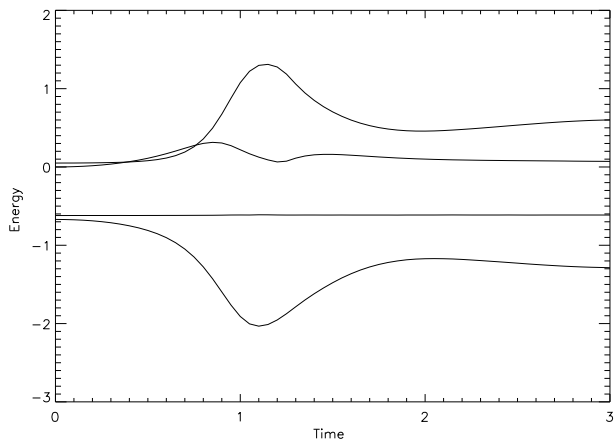,width=8.8cm}
  \caption{\label{coll_energy} Energies plotted against time for the collapse
				test. From top to bottom are the curves for the
				thermal, kinetic, total and potential energy.
				Units are given by $G = M = R = 1$.}
\end{figure}

With only 4000 particles the outgoing shock is not very sharp,
but as noted in  Steinmetz \& M{\"u}ller (\cite{Steinmetz93a}) all
global quantities are still reproduced rather well. Usually the smearing of
shock fronts is rather severe in SPH, but still this affects the evolution of
the shock fronts surprisingly little.
Depending on the problem, one often needs some factors
of 10000 particles in shock regions, to see the shocks clearly resolved.
As for the King-model test, some
conservation properties were examined and are listed in Table \ref{coll_tab}.
The shortest (individual) time-step during the simulation was $10^{-4}$,
and the particles occupied a span of 7 time-levels. In Fig. \ref{coll_energy},
the total thermal energy, total kinetic energy, total potential energy and
total energy, are plotted. The maximum thermal energy is slightly lower,
than reported by Hernquist \& Katz (\cite{Hernquist89a}) and Steinmetz \&
M{\"u}ller (\cite{Steinmetz93a}). This is because of the higher number of
neighbors used, here being 64, as compared to the 32 neighbors used by
above mentioned authors.

\subsection{Test with strong radiative cooling} \label{Merge-test}

As a test including radiative cooling, and merging of particles, we have
simulated the collapse of a rotating sphere, with parameters reminiscent
of a proto-galaxy.
This test case has been studied by Navarro \& White (\cite{Navarro93a}) and Serna et al. (\cite{Serna95a}).

The initial density field is spherically symmetric, with a radial density
profile $\rho \propto 1/r$. The sphere is started in solid body rotation
with a dimensionless spin parameter $\lambda = J|E|^{1/2}/GM_{tot}^{5/2}
\approx 0.1$ ($J$ and $E$ are the total angular momentum and
total energy). The initial radius is 100 kpc, and the total mass is
$10^{12} M_{\odot}$, with 10\% of the mass in gas and 90\% of the mass
in dark matter. The gas starts at a temperature of 1000 K.
The gas and the dark matter components are represented by 2000 particles
each.
Gravitational
softening parameters were taken to be 2 and 5 kpc for the gas and dark matter,
respectively.
  
In these simulations a significant fraction of the gas falls into unresolved
clumps in the central parts of the disk. It is therefore a demanding test
of the particle merging scheme.

\begin{figure*}
  \psfig{file=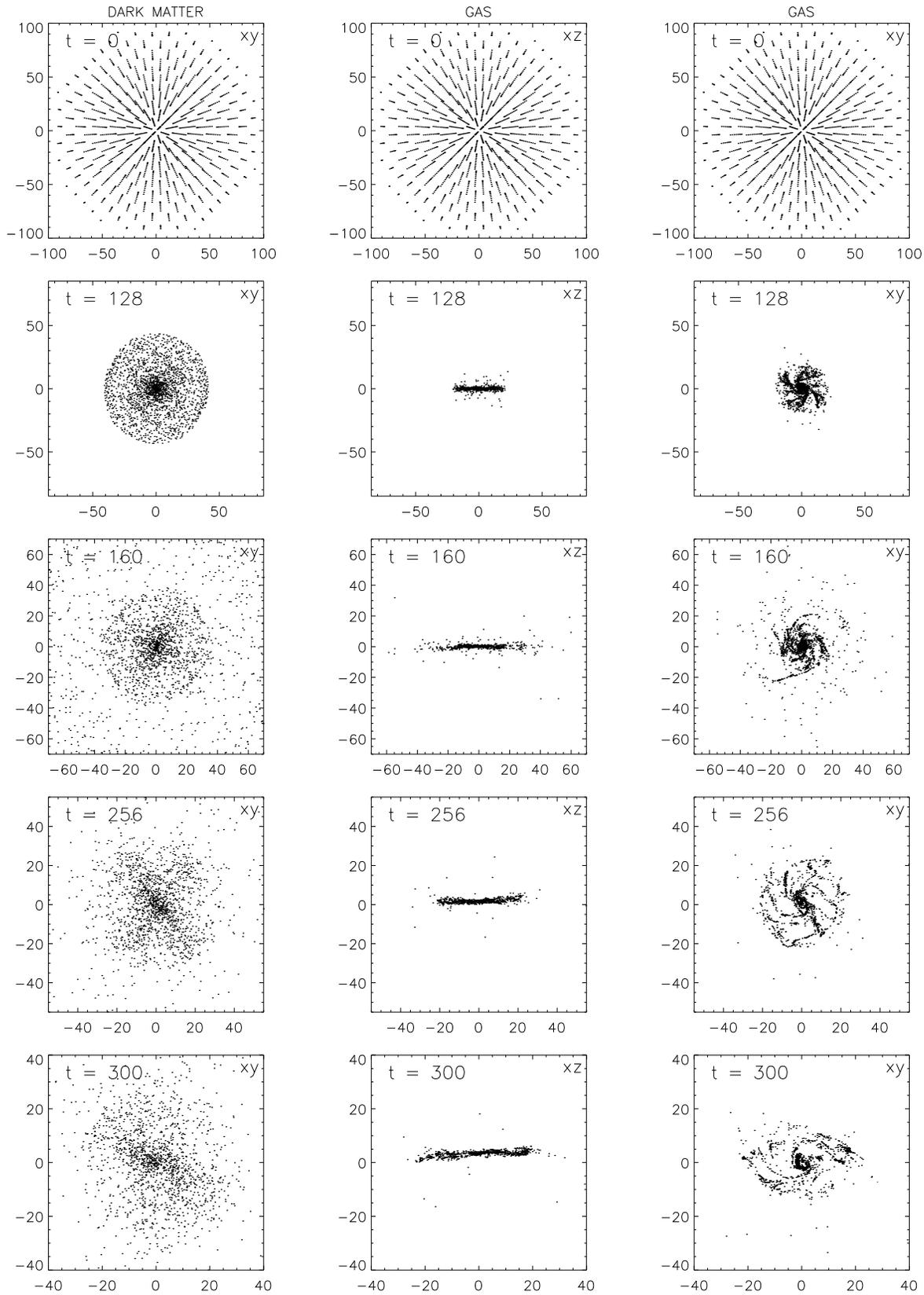,width=16cm}
  \caption{\label{NWg10} Collapse of a rotating sphere consisting of gas
  and dark matter, and with a 1/r-density profile. All numbers are given in
  the same units as Navarro \& White (\cite{Navarro93a}). (Distances are in kpc,
  and the time unit is 4.71 Myrs.) }
\end{figure*}

\begin{figure}
  \psfig{file=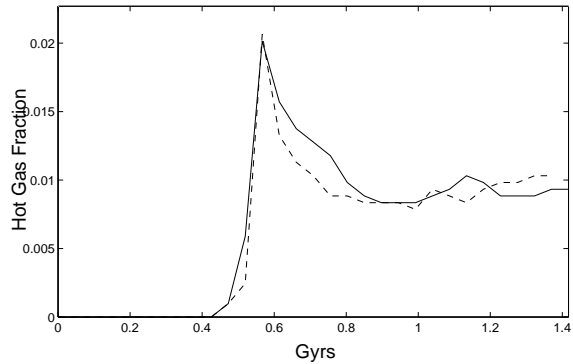,width=8cm,rheight=4.5cm}
  \caption{ \label{NWMHotFrac} The mass fraction of gas with a
  temperature above $3 \cdot 10^4$ K for the simulation,
  without merging (solid line), and with merging of particles (dashed line).}
\end{figure}

Fig. \ref{NWg10} shows the evolution of the gas and dark matter distributions
for a simulation without particle merging.
The dark matter virializes soon after the main collapse, whereas the gas
forms a thin disk. Collisional line cooling is almost
completely effective in radiating away thermal energy from adiabatic compression
and shock heating. Only a slight fraction of the gas is heated to temperatures
above 30,000 K (see Fig. \ref{NWMHotFrac}), and thermal pressure does not play any significant role in the evolution.
Radial velocities are effectively dissipated, and most of the gas has formed
a rotationally supported disk after one collapse time.

The disk displays clear spiral structures after $t \approx 160$. As reported
by Navarro \& White (\cite{Navarro93a}), the disk is unstable and starts 
breaking up into
small clumps towards the end of the simulation. Navarro \& White found that
if the gas mass fraction is reduced to 2\%, thereby
increasing the Toomre (1964) stability parameter, the disk is substantially
stabilized. Serna et al. (\cite{Serna95a}) found the stability of the disk,
when the
gas mass fraction was 10\%, to be intermediate to the 10\% and 2\% gas mass
fraction case of Navarro \& White. Our results resemble those of Serna et al.
The reason that we get a more stable disk than Navarro \& White, is probably
due to the fact that we set the SPH smoothing so as to acquire $\approx 64$
particle neighbors ($n_s = 64$), whereas Navarro \& White uses roughly
32 neighbors. When we reduce the number of SPH neighbors to $\approx 32$,
we find that the disk evolution closely matches that of Navarro \& White.

\begin{figure*}
  \psfig{file=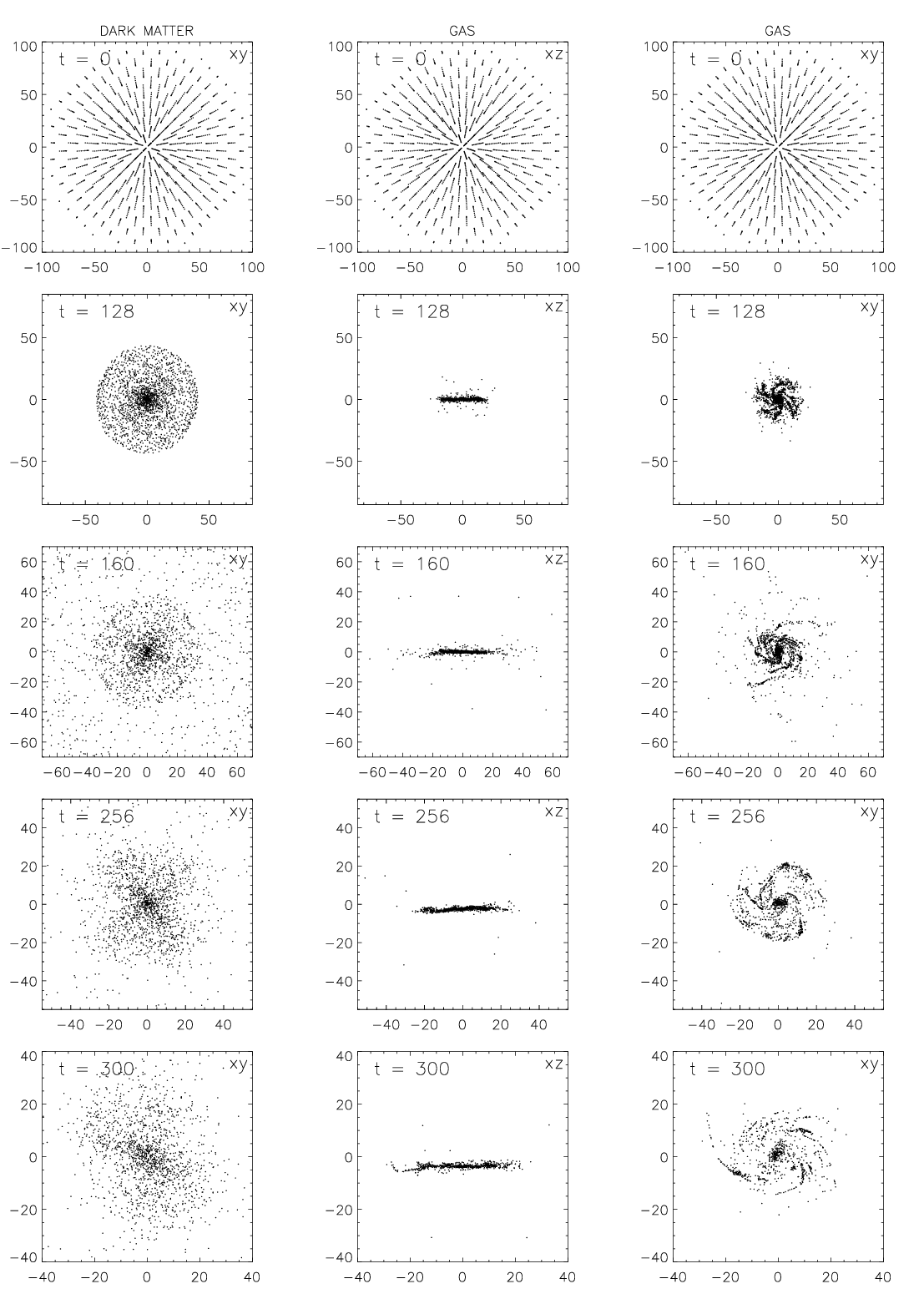,width=16cm}
  \caption{\label{NWMg10} Collapse of a rotating sphere consisting of gas
  and dark matter, and with a 1/r-density profile. All numbers are given in 
  the same units as Navarro \& White (\cite{Navarro93a}). (Distances are in kpc,
  and the time unit is 4.71 Myrs.) Particles were allowed to merge during the
  simulation. Note that individual gas particle masses vary, and therefore
  that the number density of gas particles is not proportional to the gas
  mass density. }
\end{figure*}

\begin{figure}
  \psfig{file=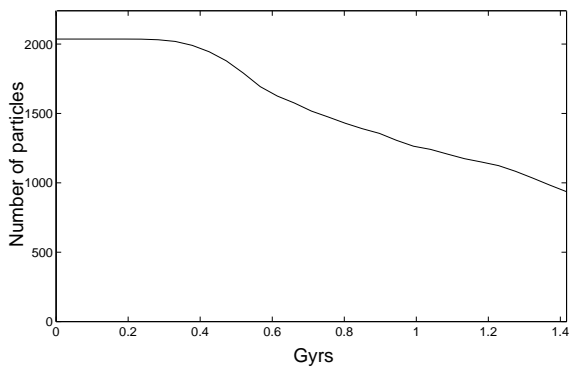,width=8cm,rheight=4.5cm}
  \caption{ \label{NWNPart} Number of gas particles, in the simulation
  with merging of particles, as a function of time.
  The number decreases as particles in high density regions merge into more
  massive particles. }
\end{figure}

In the simulation that includes our prescription for merging particles
in high density regions (see Fig. \ref{NWMg10}), the number of particles
decrease as more particles
collapse into high density regions and merge with other particles.
As can be seen from Fig. \ref{NWNPart}, the number of particles has been
reduced to half the initial value at the end of the simulation.

The gas cooling rate depends on the square of the local gas density. When
particles are merged the local resolution decreases, and small scale 
fluctuations in the density field are damped. This could potentially have
significant effects on the gas cooling. This is a fundamental problem in all
gas dynamical simulations of galaxy formation. The gas density field has to be 
assumed to be reasonably smooth on scales smaller than can be resolved in
the simulation. Fig. \ref{NWMHotFrac} shows that the fraction of hot
gas is not altered by the inclusion of particle merging.


\begin{figure}
  \psfig{file=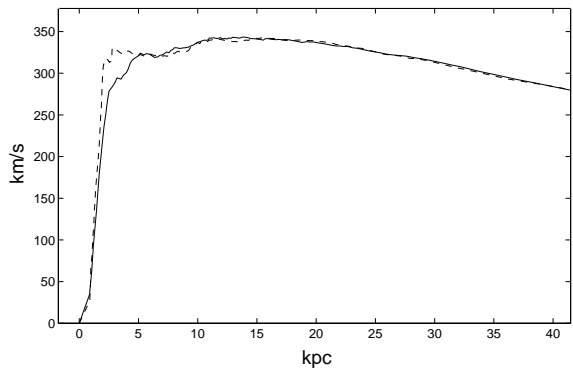,width=8cm,rheight=4.5cm}
  \caption{ \label{NWRotCurve} Circular velocity curves,
  with merging of particles (solid line), and without merging of particles
  (dashed line).}
\end{figure}

The rotation curves, ($v_c = (G M(r)/r)^{\frac{1}{2}}$), for the
simulations are shown in fig \ref{NWRotCurve}.
The mass distribution when merging is included is in excellent
agreement with the more conventional run, without merging.
The rotation curves are approximately flat out to the edge of the disk at
$\approx 25 kpc$. The disk that forms in the run without particle merging
is slightly more concentrated, with a 30\% higher gas mass inside 3 kpc.
Fig. \ref{NWg10} and \ref{NWMg10} indicates that particle merging stabilizes the
disk, and slightly suppresses the formation of substructure. This is a natural
consequence of the decreasing gas mass resolution and gravitational force
resolution, when particles merge.

The total
CPU time for the simulation is halved when particle merging is allowed,
At the end of the simulation the CPU cost per unit time has been reduced
by a factor of five, and the total
CPU time for the simulation is halved, when particle merging is allowed.

\subsection{Formation of a galactic object} \label{galax-test}

In order to test our code, and especially the particle merging scheme,
on problems that are typical of those we wish to study, we simulate the
collapse of a proto-galaxy that has been set up consistently with the
CDM cosmological model, in a fashion similar to Katz \& Gunn (\cite{Katz91c}).
The same starting conditions were used for three different simulations, two with
particle merging and one without.

\begin{center}

\begin{figure*}
  \psfig{file=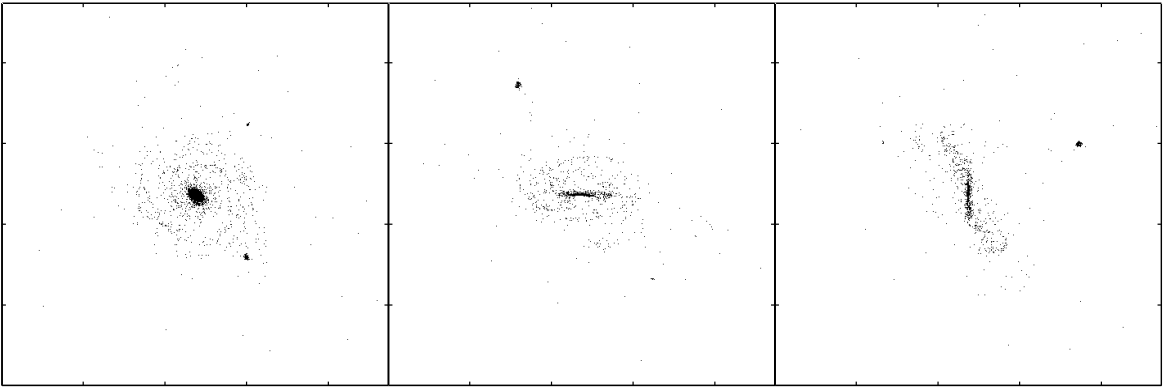,width=16cm}
  \caption{\label{GalPict8k} Three orthogonal views of projected gas particle
  positions for the simulation with 4,000 gas particles and no particle
  merging. Frames sizes are 136 kpc.}
\end{figure*}

\begin{figure*}
  \psfig{file=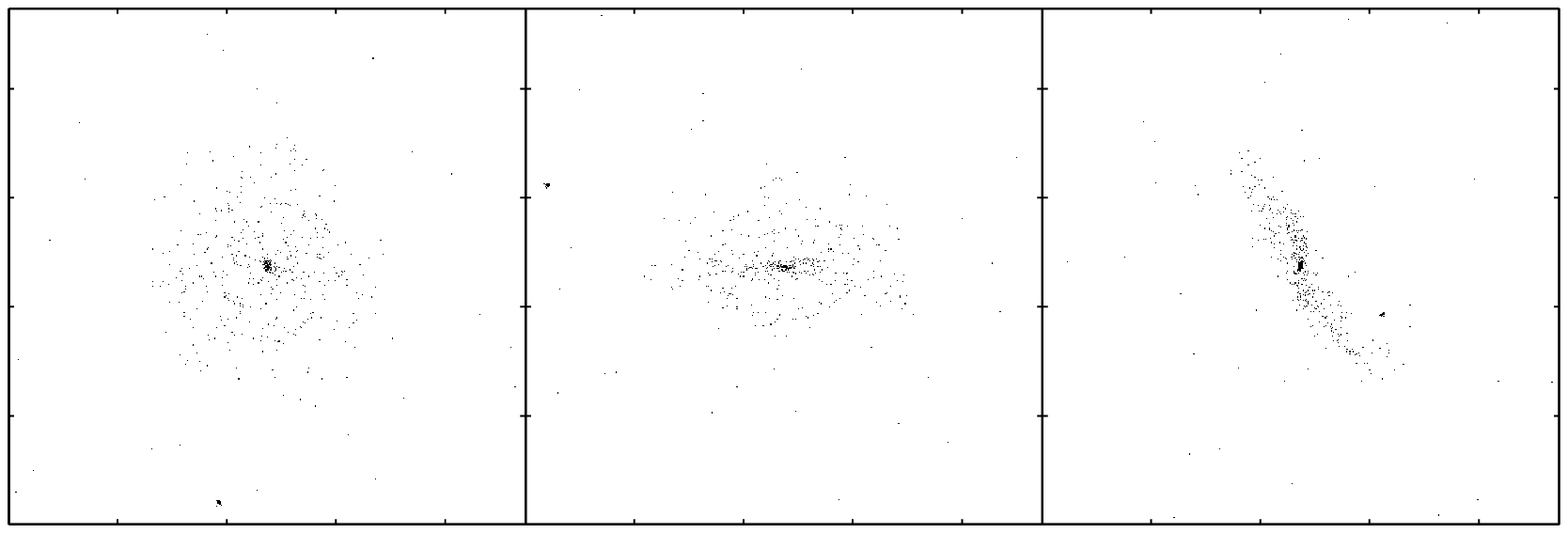,width=16cm}
  \caption{\label{GalPict8kM} Three orthogonal views of projected gas particle
  positions for the simulation with 4,000 initial gas particles and particle
  merging. Frames sizes are 136 kpc. Note that the individual particle
  mass is not constant, and that the number density of particles is therefore
  not proportional to the mass density.}
\end{figure*}

\begin{figure*}
  \psfig{file=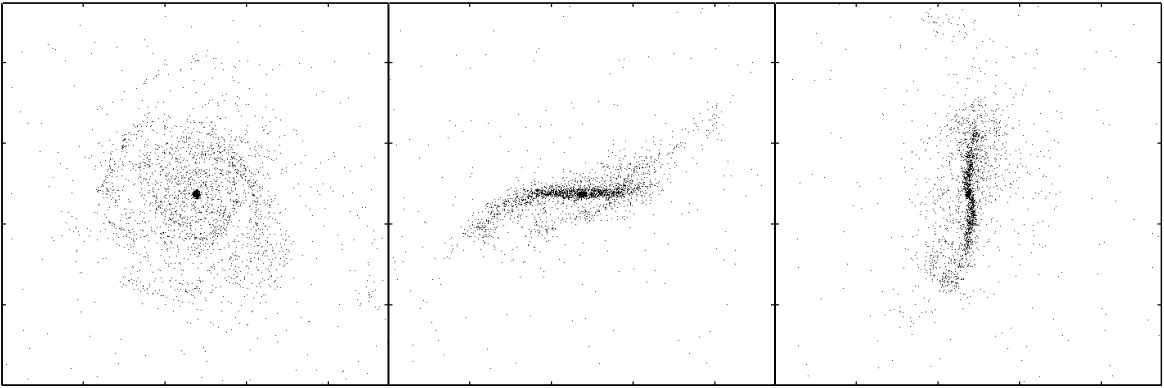,width=16cm}
  \caption{\label{GalPict45kM} Three orthogonal views of projected gas particle
  positions for the simulation with 40,000 initial gas particles and particle
  merging. Frames sizes are 136 kpc. Note that the individual particle
  mass is not constant, and that the number density of particles is therefore
  not proportional to the mass density.}
\end{figure*}

\end{center}

The Zeldovich approximation together with
a standard CDM model ($\Omega = 1$, $\Lambda = 0$, $\Omega_b = 0.05$,
$h_{100} = 0.5$, $\sigma_8 = 0.66$, Bardeen et al. \cite{BBKS86a})
power spectrum, was used to set up a cosmological
density field. The system is given an initial over-density corresponding
to a 3 $\sigma$ peak in the CDM spectrum when it is convolved with a top hat
filter of mass $10^{12} M_{\odot}$.
This density field was realized inside a sphere with a
co-moving radius of 1.46 Mpc, using both gas and dark matter
(collision-less) particles.
Gas and dark matter particles had a gravitational smoothing
of 2 and 5 kpc, respectively.
The system is started in solid body rotation, corresponding to a
dimensionless spin parameter of $\sim 0.05$, in an attempt to roughly approximate
the effects of tidal interactions. The initial gas temperature is 10,000 K,
and the gas cooling rate is that of a gas in collisional ionization
equilibrium and with a 0.1 solar metallicity,
as given by Sutherland \& Dopita (\cite{Sutherland93a}).

Three simulations were made, differing only in the number of particles used,
and whether or not particles were allowed to merge. Two simulations were
started with 8000 particles, half of them used to represent the gas and
the other half to represent the dark matter component. The only difference
between these two simulations is that one employed the previously described
scheme for merging gas particles, and the other did not. The third
simulation was made with ten times more gas particles, and with particle
merging. The same realization of initial conditions were used in all three
simulations, in order to make the final initial conditions as similar as
possible. 

The systems were started at z = 30, and evolved to z=0. The main collapse
of the proto-galaxy occurs at $z \approx 2$. At z=0 a single dominant gas
object with galactic densities has formed in all the simulations. This
galactic object consists of a compact core surrounded by a thin disk,
Fig \ref{GalPict8k}, Fig \ref{GalPict8kM}, and Fig \ref{GalPict45kM}. This
object is built up from a combination of continuous
in-fall and the merging of smaller collapsed objects.

\begin{figure}
  \psfig{file=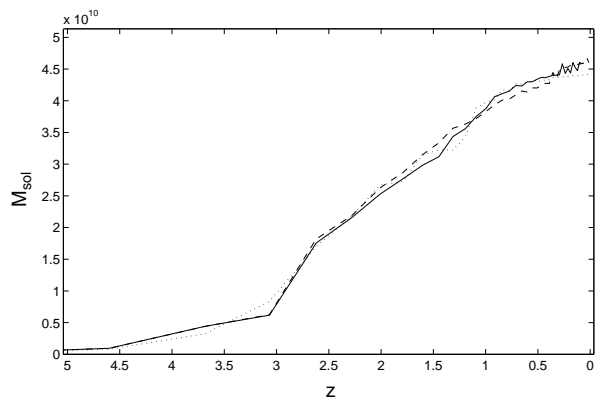,width=8cm}
  \caption{\label{ObjMassGal} The gas mass of the most massive collapsed
  object as
  a function of redshift, for 4000 particles no merging (solid line), 4000
  particles with merging (dashed line), and 40,000 particles with merging
  (dotted line).}
\end{figure}

The mass build-up of the final objects
can be seen in Fig \ref{ObjMassGal}, which shows the gas mass of the most
massive collapsed object
as a function of redshift. Collapsed gas objects were identified using a
friends-of-friends algorithm, grouping particles together that had an
over-density exceeding 1000. The magnitude of the mass build-up over time is
very similar in all three simulations.

\begin{figure}
  \psfig{file=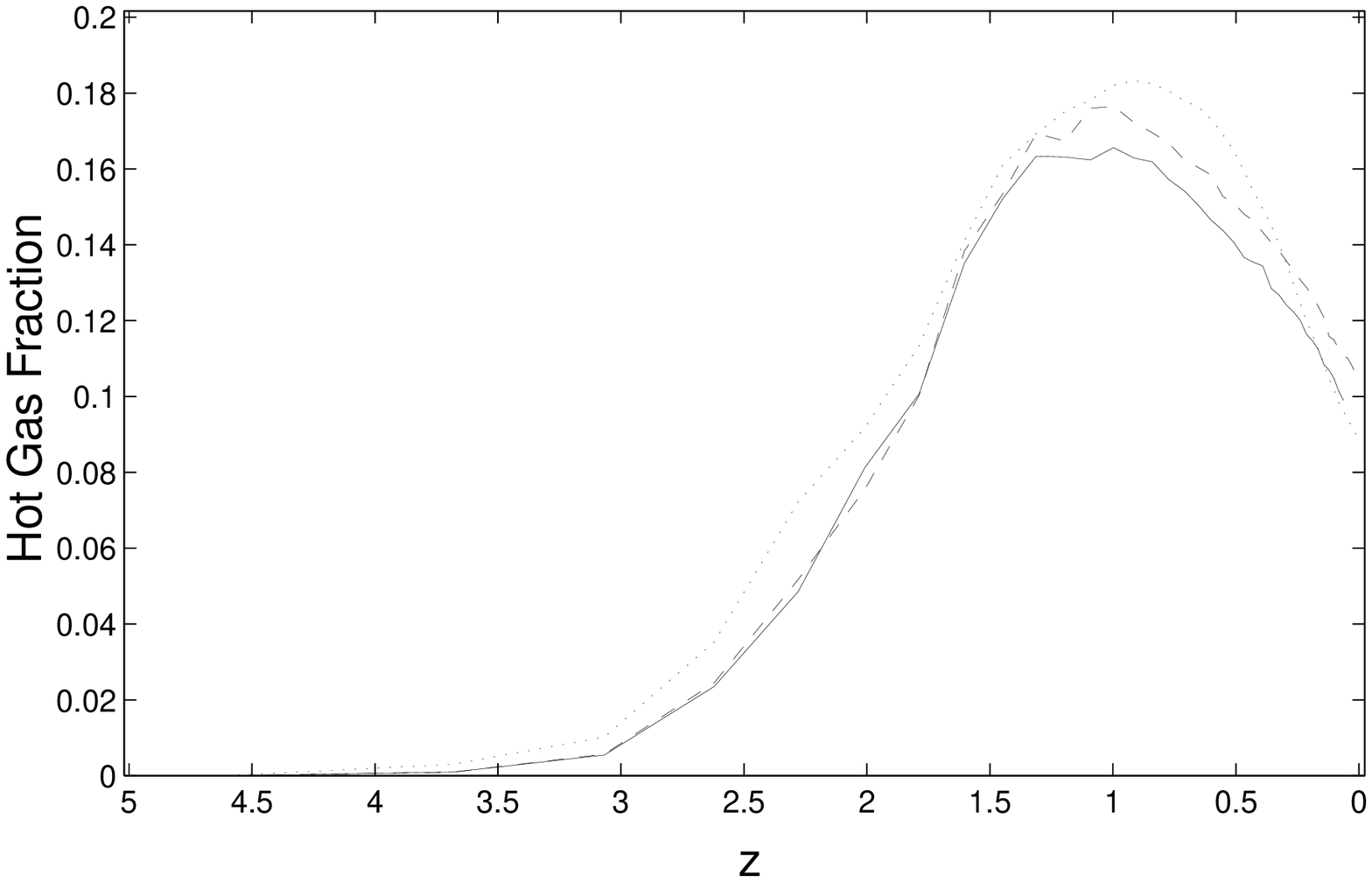,width=8cm}
  \caption{\label{HotFracGal} The mass fraction of the gas that has a 
  temperature exceeding $3\cdot 10^4 K$ as
  a function of redshift, for 4000 particles no merging (solid line), 4000
  particles with merging (dashed line), and 40,000 particles with merging
  (dotted line).}
\end{figure}

The mass fraction of gas with a temperature exceeding $3 \cdot 10^4 K$,
as a function of redshift, is shown in Fig \ref{HotFracGal}. The two
8000 particle simulations differ slightly, with  roughly 1\% more hot
gas being produced after z=1.2 in the simulation including merging.
The curve for the high resolution simulation deviates clearly from the
other two after z = 1. Between $1 < z < 0.5$ more hot gas is produced, and
after $z = 0.5$ more hot gas is able to cool than in the lower resolution
simulations. At z=0 the mass fraction of hot gas is close to 10\% for
all three simulations.

These results seem to indicate that the effects of applying the particle merging
scheme are small, outside the gravitationally unresolved cores of collapsed
gas objects. Furthermore, a ten-fold increase of the initial number of gas 
particles in a simulation produced only moderate differences,
lending some tentative support for low resolution SPH simulations of
galaxy formation.

Evolving a system with 4000 gas and 4000 dark matter particles initially,
took 64 CPU hours on a CRAY-YMP. The same system with merging of particles
required only 13 CPU hours. The high resolution system required 63 CPU
hours, almost the same as the low resolution system without merging.
The decrease in the CPU time required for a simulation, due
to particle merging, will vary with the problem and the tolerance parameters
used in the merging scheme.

\begin{acknowledgements}
The computationally demanding calculations have been performed on the
CRAY-YMP8/128, of NSC (Nationellt Superdatorcentrum), Link{\"o}ping, Sweden.
\end{acknowledgements}

\def\apj{ApJ}
\def\apjs{ApJS}
\def\mnras{MNRAS}
\def\aap{A\&A}

\end{document}